\documentclass[preprint]{aastex} 
\usepackage{amsmath}
\newcommand\unit[1]{\ \rm{#1}}

\def\cm{\unit{cm}}
\def\gm{\unit{gm}}
\def\erg{\unit{erg}}
\def\K{\unit{K}}
\def\s{\unit{s}}
\def\yr{\unit{yr}}

\def\Gy{\unit{Gy}}
\def\Ma{\unit{Ma}}
\def\Ga{\unit{Ga}}
\def\Msun{\unit{M}_\sun}
\def\Lsun{\unit{L}_\sun}

\def\tsun{\unit{t}_\sun}
\def\Mdotsun{\dot{\unit{M}}_\sun}

\begin{document} 
\title{Assessing the massive young Sun hypothesis\\ to solve the warm young Earth puzzle}
\author{David A. Minton and Renu Malhotra}
\affil{Department of Planetary Sciences, University of Arizona}
\affil{1629 East University Boulevard, Tucson, AZ 85721}
\email{daminton@lpl.arizona.edu, renu@lpl.arizona.edu}

\begin{abstract}
A moderately massive early Sun has been proposed to resolve the so-called faint early Sun paradox.  We calculate the time-evolution of the solar mass that would be required by this hypothesis, using a simple parametrized energy-balance model for Earth's climate.  Our calculations show that the solar mass loss rate would need to have been 2--3~orders of magnitude higher than present for a time on the order of $\sim2\Gy$.  Such a mass loss history is significantly at variance (both in timescale and in the magnitude of the mass loss rates) with that inferred from astronomical observations of mass loss in younger solar analogues.  While suggestive, the astronomical data cannot completely rule out the possibility that the Sun had the required mass loss history; therefore, we also examine the effects of the hypothetical historical solar mass loss on orbital dynamics in the solar system, with a view to identifying additional tests of the hypothesis.  We find that ratios of planetary orbital spacings remain unchanged, relative locations of planetary mean motion and secular resonances remain unchanged, but resonance widths and the sizes of the Hill spheres of all planets increase as the Sun loses mass. The populations and dynamics of objects near resonances with the planets as well as those of distant irregular satellites of the giant planets may contain the signature of a more massive early Sun.  However, this would not be an unambiguous test, as other dynamical processes in the early Solar System, such as the growth and presumed migration of the gas giant planets, may also be adequate to explain such signatures. 
\end{abstract} 

\keywords{Earth --- solar system: general --- solar-terrestrial relations --- stars: mass loss --- Sun: evolution}

\slugcomment{19 pages, 4 figures, 1 table; submitted to ApJ on Dec. 8, 2006}

\section{Introduction}\label{s:Introduction}
It has been well established in models of solar evolution that solar luminosity has increased over the lifetime of the Sun due to the increase in the mean density of the solar core as hydrogen is converted into helium~\citep{Gough:1981SolPhys}.  This process has led to a $\sim$30\% increase in the solar luminosity during the past $4.56\Gy$.  Because the Sun's luminosity was significantly lower in the past, calculations of Earth's ancient climate based on contemporary values of terrestrial albedo and atmospheric composition suggest that the mean surface temperature of Earth would have been below the freezing point of seawater until roughly $2\Ga$~\citep{Sagan:1972Science}.  However, there is abundant geological evidence that Earth had surface liquid water as early as $3.8\Ga$~\citep{Kasting:1989PPP}.  Geological evidence also suggests that Earth's climate may even have been warmer on average than the contemporary climate, with the exception of the relatively brief \emph{snowball Earth} epochs at $\sim2.2\Ga$, $\sim760\Ma$, and $\sim620\Ma$~\citep{Hyde:2000Nature,Hoffman:1998Science,Kirschvink:1992TPB,Evans:1997Nature}.  The problem is even more pronounced for Mars, where, even at the current levels of solar luminosity, the mean surface temperature of Mars is far too cold for liquid water to exist, and yet there is abundant evidence that Mars was warm enough for surface liquid water and that surface liquid water was present at around $3.8\Ga$~\citep{Goldspiel:1991Icarus,Jakosky:2001Nature}.

The contradiction between the cold ancient terrestrial and martian climates predicted by the lower solar luminosity and the warm ancient terrestrial and martian climates derived from geological evidence has been called the \emph{faint early Sun paradox}.  Most attempts to resolve the paradox involve models of the early atmospheres of Earth and Mars containing enhancements of atmospheric greenhouse gases.  \cite{Sagan:1972Science} proposed an early atmosphere of Earth that contained small amounts of ammonia (NH$_3$) to provide enough of a greenhouse effect to offset the lower solar luminosity.  However, it has been shown that NH$_3$ would have been rapidly dissociated by solar UV radiation, and was unlikely to be a major constituent of the early atmosphere of Earth~\citep{Kuhn:1979Icarus,Kasting:1982JGR,Pavlov:2001JGR}.  

Carbon dioxide (CO$_2$) has also been proposed as a greenhouse gas capable of resolving the paradox. A CO$_2$ concentration of at least $100\times$ present atmospheric level (PAL) would have to have been present to prevent the Archean Earth from freezing over~\citep{Kasting:1987PreCambRes}.  According to \cite{Rye:1995Nature}, this abundance of CO$_2$ in the atmosphere should have led to the common presence the mineral siderite (FeCO$_3$) in Archean paleosoils, but the absence of siderite from known soils older than $2.2\Ga$ suggests that CO$_2$ could not have been present in the Earth's atmosphere in the required amount.  Analysis of the weathering rinds of river gravels dated to $3.2\Ga$ suggest that the presence of Fe(II)-rich carbonate in the rinds sets a lower limit of CO$_2$ partial pressure in the atmosphere to only several times the present value, which is two orders of magnitude below what is required to keep the surface temperature of the Earth above freezing~\citep{Hessler:2004Nature}.  More recently, \cite{Ohmoto:2004Nature} have challenged the interpretation of Rye et al. regarding the lack of siderite in paleosoils, arguing that if even a very small amount of oxygen was present in the Archean Earth's atmosphere, siderite (or any other ferrous-rich mineral) would have been unstable, and CO$_2$ could still have been present in high enough concentration to warm the Earth.

Methane (CH$_4$) has also been proposed as a possible greenhouse gas responsible for keeping the Archean Earth warm~\citep{Hart:1978Icarus,Kiehl:1987JGR}.  Although CH$_4$, like NH$_4$, is susceptible to photodissociation by solar UV, it can remain in the atmosphere for much longer and methanogenic bacteria could maintain the required atmospheric levels~\citep{Pavlov:2000JGR}.  Were there adequate populations of methanogenic bacteria available in the early Archean? With only trace amounts of evidence of life even existing prior to about $3.6\Ga$, it is difficult to support the evidence of a warm early Earth throughout the Hadean and early Archean with methanogenic bacteria replenishing atmospheric CH$_4$.  

For Mars, one long standing problem with solving the faint young sun paradox with a dense CO$_2$ atmosphere is that models predict that the CO$_2$ will begin to condense out as a cloud layer at the required atmospheric pressures, thereby increasing the global albedo and off-setting the warming greenhouse effect~\citep{Kasting:1991Icarus}.  \cite{Pierrehumbert:1998JAS} proposed a model of the early martian atmosphere in which CO$_2$ cloud particles of a certain size could cause a scattering greenhouse, requiring an atmospheric pressure of less than 1~bar CO$_2$ to keep early Mars above the freezing point of water.  However, laboratory studies of CO$_2$ cloud formation under martian conditions seem to suggest that the types of clouds that could form on Mars, even with a CO$_2$ atmosphere with pressures as high as 5~bars would not warm the planet above the freezing point of water~\citep{Glandorf:2002Icarus,Colaprete:2003JGR}. 

An alternative solution to the faint early Sun paradox involves a non-standard solar model in which the Sun has lost significant mass over time.  A more massive early Sun will have two effects.  First, since stellar luminosity is mass-dependent, a larger solar mass implies a correspondingly larger solar energy output.  Second, owing to the existence of adiabatic invariants of the keplerian orbits, the planets would have orbited closer to the Sun had the solar mass been higher.  There is an upper limit to how much more massive the Sun could have been in the past based on the constraint that if the solar flux at Earth had been $\sim10$\% higher at any time in the past then the Earth would have lost its water due to a moist greenhouse atmosphere, in which water reaches the stratosphere and is lost due to UV dissociation and escape of hydrogen~\citep{Kasting:1988Icarus}. Such a process is thought to have occurred on Venus~\citep{Kasting:1983Icarus}.

\cite{Whitmire:1995JGR} suggested that a young Sun 3\%--7\% more massive would be able to explain the evidence of liquid water on the surface of Mars at $3.8\Ga$.  \cite{Sackmann:2003ApJ} tested solar models in which the Sun first entered the main sequence with an initial mass of up to $1.07\Msun$.  Using various mass loss rate functions that were constrained such that after $4.56\Gy$ on the main sequence the solar models had the currently observed solar wind mass loss rate and the current mass for the sun, they compared the resulting solar interiors of their models against solar interior profiles (adiabatic sound speed and density) based on contemporary helioseismic measurements. They concluded that the mass losing Sun models were consistent with helioseismic measurements, with the 7\% more massive sun case marginally more consistent than the standard solar model.  While promising, the technique used by Sackmann \& Boothroyd can neither support nor rule out a solar model with a mass loss of about 7\% until improvements are made in helioseimic observational sensitivity and models of solar input physics.

The Sun currently loses a small amount of mass due to coronal mass ejections ($\sim10^{-15}\Msun\yr^{-1}$), the solar wind ($\sim2$--$3\times 10^{-14}\Msun\yr^{-1}$) and radiation ($\sim7\times 10^{-14}\Msun\yr^{-1}$), the latter due to conversion of matter into energy through thermonuclear fusion at the core~\citep{Feldman:1977SOIV,Shu:1982TPU}.  Assuming that each of these rates have remained constant over time, the total amount of mass lost by the Sun over its $4.56\Ga$ history is only about 0.05\% of the total solar mass.  However, the physical processes that set the rate of mass loss due to solar wind and coronal mass ejections are poorly understood and it is not unreasonable to explore solar models in which the rates of mass loss were larger in the past by orders of magnitude.  Astronomical observations indicate generally larger mass loss rates by stellar winds for younger solar analogues~\citep{Wood:2002ApJ,Wood:2005ApJ}.  There is some evidence in lunar regolith samples of a long term secular decrease in solar wind flux during the past $3\Gy$~\citep{Kerridge:1991TSIT}.  There is also evidence in the meteoritic record of enhanced solar activity during the solar system's early history, though this has been interpreted as evidence of the Sun's very active but brief T~Tauri phase, rather than an extended period of enhanced solar wind~\citep{Caffee:1987ApJ}.  

In this paper we examine anew the hypothesis of a mass losing Sun.  First we constrain the solar mass loss rate time history that would be required to solve the faint early Sun paradox by requiring that Earth's mean surface temperature remain above $273\K$ for all of its history (Section~\ref{s:Climate}).  We then compare the resulting mass loss functions with estimates in the literature for the stellar wind mass loss rates of younger sun-like stars (Section~\ref{s:Winds}).  The comparison is discouraging:  the astronomical data does not support a solar mass loss history that would resolve the faint early Sun paradox.  However, astronomical data cannot rule out the hypothesis either. Therefore, with a view towards identifying other tests of the hypothesis, we explore the effects of a mass losing Sun on the orbital dynamics of the solar system, including how a 7\% more massive Sun would affect mean motion and secular resonances, the irregular satellites of the Jovian planets, and other small bodies in the solar system (Section~\ref{s:Dynamics}).  We summarize our results and conclusions in Section~\ref{s:Conclusion}.
   
\section{Effect of Solar Mass on Climate History}\label{s:Climate}
A very simple climate model can be used to explore the dependence on the mass of the Sun of the mean surface temperature of the Earth, while keeping all other factors constant.  The steady state mean surface temperature of the Earth (or any other terrestrial planet) is given by the following energy balance equation~\citep{Pollack:1979Icarus-a}: 
\begin{equation} 
   \left(1-A\right)S\pi R^2=\sigma\varepsilon T_s^4 4\pi R^2 
   \label{e:pollack-MST} 
\end{equation} 
where $A$ is the average planetary albedo, $S$ is the solar flux at the top of the atmosphere, $R$ is the planetary radius, $\varepsilon$ is the atmospheric IR emissivity, and $\sigma$ is the Stefan-Boltzmann constant of $\sigma=5.67\times 10^{-5}\gm\cdot\s^{-3}\cdot\K^{-4}$.   Typical values for the Earth are $A=0.34$ and $\varepsilon=0.6$~\citep{Pollack:1979Icarus-a,Hartmann:1994GPC}.  For the purposes of this paper, we make the minimalist assumption that these parameters are time-independent and we take the current typical values to be constant over all of geologic time.  This is not necessarily the most accurate way to model the Earth's atmosphere, since changes in atmospheric greenhouse gas composition can vary the emissivity parameter $\varepsilon$ and changes to surface composition and cloud cover can change the planetary albedo parameter $A$; this parametrization may also be criticized as insufficient to represent the variety of greenhouse effects that may be possible.  However our purpose is to identify solar mass histories that simply and completely resolve the faint early Sun paradox without the need to invoke other effects.  Thus, for our purposes Equation~\ref{e:pollack-MST} provides an adequate representation, and can be rearranged to give the terrestrial surface temperature as a function of solar flux: 
\begin{equation} 
   T_s=\left[\frac{S\left(1-A\right)}{4\varepsilon\sigma}\right]^{1/4}.
   \label{e:MST} 
\end{equation} 
 
\citet{Gough:1981SolPhys} reports that, based on stellar nucleosynthesis and stellar evolution models, the rate of increase in luminosity of the Sun with time, assuming the solar mass is constant, can be represented as
\begin{equation} 
   L(t)=\left[1+\frac{2}{5}\left(1-t/\tsun\right)\right]^{-1}\Lsun,
   \label{e:Lvst} 
\end{equation} 
where $\tsun=4.56\times 10^9\yr$ is the current age of the Sun \citep{Dearborn:1991TSIT}, and $\Lsun\approx 3.9\times 10^{33}\erg\sec^{-1}$ is the current solar luminosity~\citep{Shu:1982TPU}.
We will assume that the Earth's orbital eccentricity can be neglected and that the Earth's orbital radius is equal to its semi-major axis, $a$. 
 
\begin{figure}
\plotone{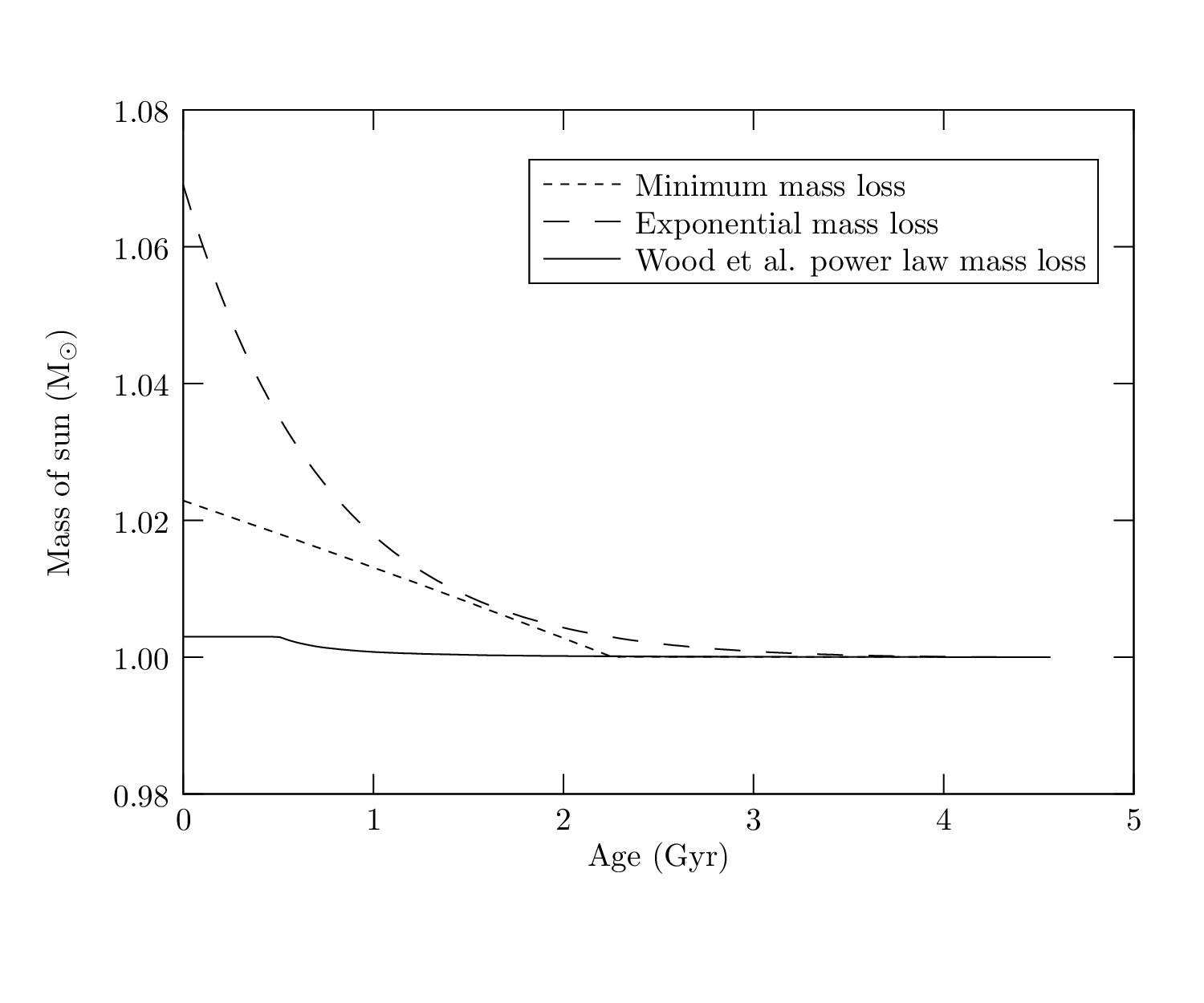}
\caption{Solar mass over time for the three different solar mass history models considered.\label{f:solarmass}}
\end{figure}

A mass losing Sun will affect the Earth's climate in two ways.  First, the main sequence stellar luminosity is quite sensitive to stellar mass, $L\propto M^P$, where $P$ is in the range 3--5 depending upon sources of opacity~\citep{Iben:1967ARAA}; for the early Sun, we adopt $P=4.75$, following \citet{Whitmire:1995JGR}.  Second, if the solar mass loss is slow compared with the orbital motion of the planets, then it follows from the adiabatic invariance of the actions for a keplerian orbit that
\begin{equation}
   \left[M(t)+M_i\right]a_i(t)=\textrm{constant},
	\label{e:jeansa}
\end{equation}
where $M_i$ and $a_i$ are the mass and orbital semimajor axis of a planet (or minor planet) in the solar system, and $M(t)$ is the time-varying solar mass. 
In the solar system, $M_i\ll\Msun$, therefore a planet's semi-major axis will be smaller for a more massive Sun by $a_i\propto M^{-1}$.  The solar radiation flux at Earth, $S$, is related to the Sun's luminosity $L$ and Earth's semi-major axis $a$ by $S\propto La^{-2}$.  Therefore, the time-variation of the solar flux at Earth is given by
\begin{equation}
	S(t)=S_0\left[1+\frac{2}{5}\left(1-t/\tsun\right)\right]^{-1}\left(\frac{M(t)}{\Msun}\right)^{6.75},
	\label{e:SvstvarM}
\end{equation}
where $S_0$ is the current solar radiation flux at the top of Earth's atmosphere, $S_0=1.37\times 10^6\erg\cdot\cm^{-2}\cdot\s^{-1}$. 
Using Eqnuation~(\ref{e:SvstvarM}) in Equation~(\ref{e:MST}), we obtain the time dependence of the terrestrial surface temperature:
\begin{equation} 
   T_s\left(t\right)=\left\{\frac{S_0\left(1-A\right)} 
   {4\varepsilon\sigma\left[1+\frac{2}{5}\left(1-t/\tsun\right)\right]}\right\}^{1/4}\left(\frac{M(t)}{\Msun}\right)^{1.69}.
   \label{e:Tvst} 
\end{equation} 
Note that Equation~\ref{e:Tvst} is obtained from a straightforward energy balance model.  We now use it to construct the ``minimum mass-loss'' history for the Sun that keeps the terrestrial surface temperature $T_s(t)$ above $273\K$ for all time, and is also consistent with the current solar mass loss rate of $\Mdotsun\approx -2\times10^{-14}\Msun\yr^{-1}$.  
We do this as follows.  We assume that the current solar mass loss rate has remained constant for the past $2.3\Gy$, the time span over which the standard solar model yields terrestrial surface temperature above $273\K$~\citep{Sagan:1972Science}.  For times prior to $2.3\Ga$, we calculate the solar mass that would be required to just keep the terrestrial surface temperature at $273\K$.  Therefore, we specify the terrestrial surface temperature at $t=0$, $T_s(0)=273\K$.  Equation~(\ref{e:Tvst}) then yields the initial solar mass, $M(0)=1.026\Msun$.  Also from Equation~(\ref{e:Tvst}), we obtain the solar mass as a function of time that keeps the terrestrial surface temperature constant, $T_s(t)=T_s(0)=273\K$: 
\begin{equation}
M(t)=0.973\Msun\left[1+\frac{2}{5}\left(1-t/\tsun\right)\right]^{0.148};
\label{e:ConstTempM}
\end{equation}
the corresponding mass loss rate is given by
\begin{equation}
\dot{M}(t)=-1.26\times10^{-11}\left[1+\frac{2}{5}\left(1-t/\tsun\right)\right]^{-0.852} \Msun\mbox{ y}^{-1}.
\label{e:ConstTempMdot}
\end{equation}
In equations (\ref{e:Tvst}) and (\ref{e:ConstTempMdot}) we have adopted the numerical values of $A$ and $\varepsilon$ typical of Earth mentioned above, and $\tsun=4.56\times10^9\yr$.  By demanding continuity of the solar mass time variation and consistency with the present solar mass loss rate, we obtain the piecewise ``minimum solar mass loss'' model:
\begin{equation}
M(t)=
\begin{cases}
	0.974\Msun\left[1+\frac{2}{5}\left(1-t/\tsun\right)\right]^{0.15} & \text{for $t\leq 2.39$ Gy,} \\
	\Msun +\Mdotsun(t-\tsun) & \text{for $t> 2.39$ Gy.}
\end{cases} 
\end{equation}
This function is shown by the short dashed curve in Figure~\ref{f:solarmass}.  

Beyond the minimum solar mass loss history, we consider a solar mass loss rate that has decreased exponentially with time.  We construct this with parameters constrained to keep the mean surface temperature of Earth above the freezing point of water for all of the planet's history.  A functional form of this mass loss rate can be specified as $\dot M(t) \sim c + d(e^{-\alpha t}-e^{-\alpha\tsun})$, which implies solar mass as a function of time as follows:
\begin{equation}
M(t)=\Msun+C(t-\tsun)+D(e^{-\alpha t}-e^{-\alpha\tsun}).
\label{e:expmsun}
\end{equation}
If we denote the initial solar mass, $M(0)=m_f\Msun$, then the constants $C$ and $D$ are as follows:
\begin{eqnarray*}
C & = & \Mdotsun+\alpha De^{-\alpha\tsun}, \\
D &=& \frac{(m_f-1)\Msun+\Mdotsun\tsun}{1-e^{-\alpha\tsun} },
\end{eqnarray*}
where $\Mdotsun=\dot M(\tsun) \simeq -2\times10^{-14}\Msun\mbox{ y}^{-1}$ is the present solar wind mass loss rate (at $t=\tsun$).
The time constant, $\alpha$, is found by considering the requirement that the mean surface temperature of Earth remained above the freezing point of water during all of Earth's history:  $T\left(t<\tsun\right)>273\K$.  This yields $\alpha=1.32\times 10^{-9}\yr^{-1}$.  The parameter $m_f$ is a free parameter, except that, as discussed above, \citet{Kasting:1988Icarus} argues that the solar flux cannot have been more than 10\% higher at any point in the past since Earth has not lost its oceans due to a runaway moist greenhouse atmosphere.  This provides an upper limit on the initial mass factor of $m_f=1.07$, and \citet{Sackmann:2003ApJ} calculate that this value is consistent with helioseismology.  With this value, we have $C\simeq-9.2\times10^{-11}\Msun\mbox{y}^{-1}$ and $D\simeq0.070$.  The resulting solar mass function $M(t)$ is shown in Figure~\ref{f:solarmass}.

\begin{figure}
\plotone{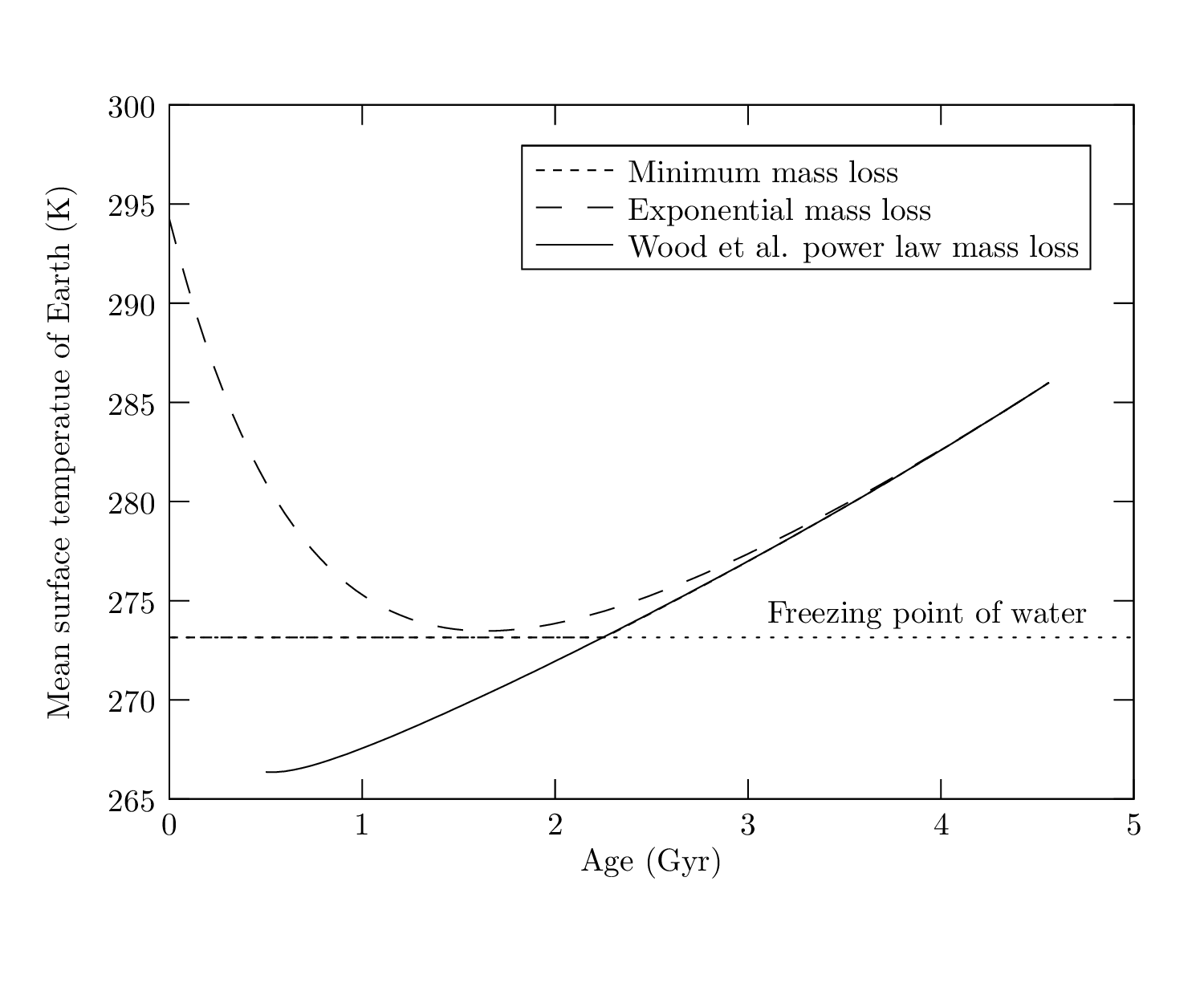}
\caption{Estimated mean surface temperature of the Earth calculated using Equation~\ref{e:Tvst} for the three different solar mass history models considered.\label{f:earthtemp}} 
\end{figure}

For the mass loss functions considered above, the corresponding mean surface temperature of Earth as a function of time is plotted in Figure~\ref{f:earthtemp}.  Note that although the cumulative mass loss of the Sun required to resolve the faint early sun paradox is modest, the timescale required for this mass loss is quite long, ${\cal O}(10^9)$\yr.

We briefly consider Mars.  With the standard solar model, in order to keep the global surface temperature on Mars above $273\K$ prior to $3.5\Ga$, the effective atmospheric emissivity would need to be $\varepsilon=0.29$ (assuming its present albedo, $A=0.16$), or $\varepsilon=0.23$ (assuming an Earth-like albedo, $A=0.34$).  For the mass losing solar models considered here, for the minimum required $\dot{M}$ that solves the faint early sun problem for Earth, in order to solve the problem for Mars the effective emissivity of Mars would need to be $\varepsilon=0.33$ (assuming its present albedo of $A=0.16$) or $\varepsilon=0.26$ (assuming an Earth-like albedo, $A=0.34$). 

\begin{figure}
\plotone{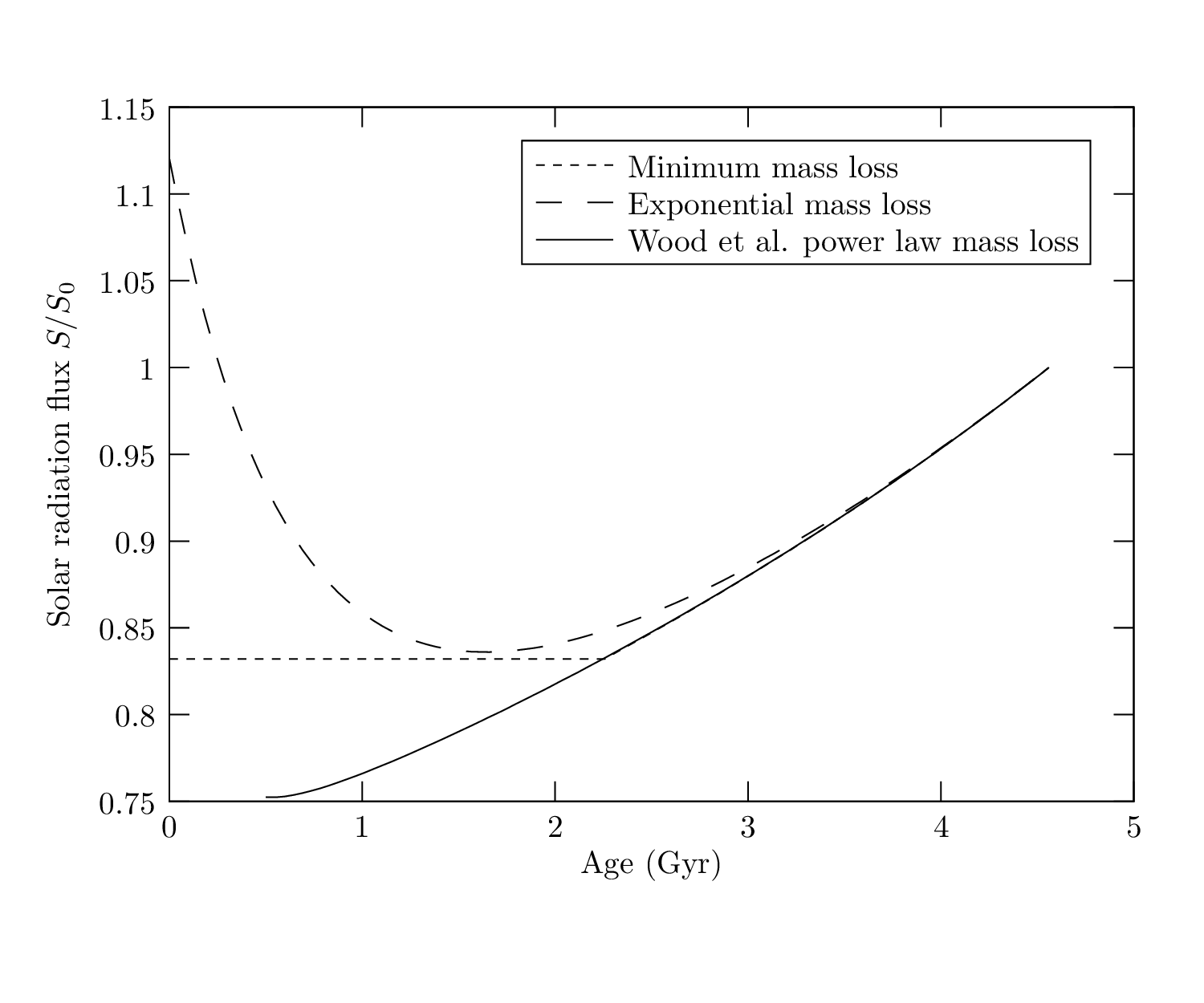}
\caption{Solar radiation flux, $S(t)$, for any planet relative to the current solar radiation flux, $S_0$.\label{f:solarrad}}
\end{figure}

\cite{Kasting:1991Icarus} calculated what the required solar luminosities to keep Mars above the freezing point of water with a massive CO$_2$ atmosphere and including the effects of CO$_2$ condensation.  He found that in order to keep early Mars warm enough for liquid water with a massive CO$_2$ atmosphere, the ratio of the solar radiation flux to the current solar radiation flux ($S/S_0$) had to be greater than 0.80--0.86.  If the solar flux ratio was lower than this value, then the decrease in the convective lapse rate in the troposphere of Mars due to the condensation of CO$_2$ clouds would offset the greenhouse warming enough that the surface could never be above the freezing point of water, no matter how high the CO$_2$ pressure was.  Figure~\ref{f:solarrad} shows the solar radiation flux ratio as a function of time for the solar mass loss models considered here.  Both the models considered here that solve the faint warm early Earth problem are also capable of satisfying Kasting's criteria for keeping early Mars warm. 

\section{Stellar Winds of Sun-like Stars}\label{s:Winds}

Since the sun is thought to be a typical G type dwarf star, it would be informative to measure the stellar wind mass outflows of other sun-like stars at various stages of their evolution on the main sequence.  Stellar winds from G and K type dwarf stars are very difficult to observe directly owing to their low optical depth.  Recently however, indirect measurements of the stellar winds of a small number of nearby sun-like stars have been made using the {\em Hubble Space Telescope}.  These measurements exploit the charge exchange that occurs when the ionized stellar wind collides with the neutral interstellar medium, and which is detectable in a H\textsc{i} Ly$\alpha$ absorption feature~\citep{Zank:1999SSR,Wood:2002ApJ}.  By measuring amount of Ly$\alpha$ absorption in a star, and fitting the observed absorption feature to a model astrosphere interacting with its local interstellar medium, the mass loss rate of the star due to stellar winds can be estimated.  The density and relative velocity of the local interstellar medium with respect to the star must also be known.  

Using this method, \citet{Wood:2005ApJ} have estimated the stellar wind fluxes of several nearby solar-type main sequence stars.  For each star, Wood et al. reported a mass loss rate due to stellar winds relative to the currently measured solar wind mass loss rate.  They also reported the stellar x-ray luminosity, and the stellar surface area.  They then used a power law relationship between the stellar x-ray flux and the age as reported by \citet{Ayres:1997JGR},
\begin{equation}
   F_x\propto t^{-1.74\pm 0.34}.
	\label{e:XrayAge}
\end{equation}
We use Equation~\ref{e:XrayAge}, together with the Sun's age as a reference point, to obtain an age estimate for each of the stars in Wood et al.'s sample.    We also calculated the errors in the stellar ages using the error in the power law relationship between stellar x-ray flux and age.  In addition, we also searched the literature for independent measurements of stellar ages, if available.  The ages estimated from Equation~\ref{e:XrayAge} do not always agree with the ages found by other methods, and some are well outside the error bars of the x-ray flux ages. We consider here a sub-set of Wood et al's stars, i.e., only the eleven G and K stars for which mass loss rates were obtained.  This data is listed in Table~\ref{t:star_data} and plotted in Figure~\ref{f:solarmassrate}.  Mass loss function that fit to the data of Wood et al. with a power law are also plotted in Figure~\ref{f:solarmassrate}, along with the two solar mass loss history models described in Section~\ref{s:Climate}.  

It is clear that the mass loss rate functions that we derived in the previous section to solve the faint early Sun paradox are inconsistent with the astronomical data on the mass loss rates of sun-like stars as found by Wood et al. The required solar mass loss rates at early times (0--2~Ga) are about 1--2~orders of magnitude higher than that for the observed solar analogues.  The cumulative total stellar mass loss inferred from the Wood et al. data is $\lesssim 0.003\Msun$, in contrast with the required 0.03--0.07$\Msun$.  The data also indicates a timescale of ${\cal O}(10^8)\yr$ for the decrease of stellar wind flux at early times, whereas our hypothesis requires a slower decay, on timescale $\sim10^9\yr$.  
To summarize, the faint young sun paradox requires a cumulative solar mass loss and early solar mass loss rates that are about 1--2~orders of magnitude and about one order of magnitude, respectively, higher than the astronomical data have revealed.  The caveat is that the stellar wind flux estimates of the younger solar analogs are model-dependent.  \cite{Wood:2002ApJ} estimate that there may be a factor of two uncertainty in their estimated mass loss rates due to the uncertainty in the stellar wind velocity.  

\section{Dynamical Effects of Higher Mass Sun}\label{s:Dynamics}
In the above analysis we have shown that in order for a mass losing Sun to solve the faint early Sun paradox, the Sun's mass had to be significantly higher for a period of time on the order of $2\Gy$.  In this section, we explore what effects such a higher mass Sun might have had on the dynamics of the solar system, and whether some signatures of the early massive Sun may remain imprinted on the orbital dynamics of the present planetary system.  A non-exclusive list of dynamical effects is discussed below.  

\subsection{Planetary Orbits}\label{ss:plorbits}
A slow solar mass loss conserves all the three actions of a keplerian orbit.  Therefore, orbital eccentricities and inclinations of all planetary orbits are unchanged by a solar mass loss, and it was shown above that the semimajor axes of each of the planets are related to the time-varying mass of the sun through Equation~\ref{e:jeansa}.  This means that, for an initial mass of the Sun greater than present by a factor $m_f$, the net expansion of all planetary orbits has been a factor $m_f$ relative to the initial orbits, and the orbital periods $T_i$ have increased by a factor $m_f^{2}$.  Thus, for $m_f\approx 1.07$, and considering the mass loss histories described in Section~\ref{s:Climate}, the year would have been shorter by 2--6\% during the Archean epoch (Earth age 0.5--2.5$\Gy$, see Figure~\ref{f:solarmass}).  Indicators of the length of the year in the geological record (tidal rhythmites, banded iron formations) provide poor precision for pre-Cambrian epochs~\citep{Williams:2000RvGeo,Eriksson:2000Geo}, and are unlikely to provide a test of a shorter year .

The slightly more compact ancient planetary system would have experienced relatively stronger orbital perturbations arising from the mutual gravitational interactions of the planets.  The rates of secular precession of apsides and of orbit poles, which are proportional to $M_j/T_i\Msun$, would have been faster by a factor $m_f$.  Although these precessional motions have been linked to climate cycles on timescales of ${\cal O}(10^5$--$10^6)$ years in recent Earth history~\citep{Hinnov:2000AREPS}, the much lower resolution of the very ancient climate record~\citep{Erwin:2006AREPS} makes it unlikely that it could be useful in testing these implications of the higher ancient solar mass.

\subsection{Minor Planet Resonance Locations}\label{ss:Resonance}
Locations of mean motion resonances depend on the ratios of the semimajor axes between the planets and a test particle. It was previously shown that the semimajor axes of the planets are related to the time varying mass of the sun through Equation~\ref{e:jeansa}.  For a Sun with a higher mass given by the factor $m_f$, the change in the ratio of the semimajor axis of Jupiter to a massless particle can be found by:
\begin{equation}
   \frac{a_p/a_J}{a_p'/a_J'}=\frac{m_f\left(M_\sun+M_J\right)}{m_f M_\sun+M_J}
	\label{e:aratio}
\end{equation}
Where the primed values refer to the pre-mass loss semimajor axes.  For a mass loss of 7\%, this gives a change in the ratio of semimajor axes of
$$\frac{a_p/a_J}{a_p'/a_J'}=1.00006$$
The assumption that $a_i\propto M_\sun^{-1}$ is therefore a good one when considering the locations of mean motion resonances.  It follows that the relative locations of mean motion resonances of the major planets remain nearly unchanged as the solar mass changes, however the strengths of resonant perturbations are not invariant: the fractional amplitude of resonant perturbations is proportional to $\sim (M_i/\Msun)^{1/2}$, so that resonance widths (measured as a range in semimajor axis), $\Delta a$, scale as~$\sim m_f^{-3/2}$. The Kirkwood Gaps and the Hilda group of asteroids in the main asteroid belt may have been subjected to the effects of an expanding resonance width due to solar mass loss.  For the Kirkwood Gaps, asteroids from near the edges of the resonances would be destablized, but this would not leave a trace of the original boundaries of the gaps in the present asteroid orbital distribution.  For the surviving Hilda asteroids, the result would be a more compact final orbital element distribution compared to their initial orbits; the net changes estimated are small, not inconsistent with the observed population, but also consistent with other dynamical processes in the early solar system~\citep[e.g., resonance sweeping that accompanied the orbital migration of the planets][]{Franklin:2004AJ}.

One of the strongest resonant perturbations on asteroids is the $\nu_6$ secular resonance which defines the inner edge of the present-day asteroid belt.  Considering the effects of the time varying solar mass, we find that the free precession rates of minor planets as well as the secular frequencies of the major planets are both proportional to $m_f$, therefore the locations of secular resonances are invariant relative to the orbits of the major planets. Thus, we expect no signature of the solar mass loss in the inner boundary of the asteroid belt. 

\subsection{Irregular satellites of the outer planets}\label{ss:Hill}

The irregular satellites of the outer giant planets are small objects having semimajor axes, eccentricities, and inclinations much higher than the regular satellites; the majority of them are on retrograde orbits. They are thought to be objects that formed in heliocentric orbits which were subsequently captured into stable orbits around the giant planets.

The irregular satellites all orbit well within their planet's Hill sphere.  In the restricted three body problem, the Hill sphere is defined by the radius
\begin{equation}
   r_H=a_2\left(\frac{Gm_2}{3m_c}\right)^{1/3}
        \label{e:HillR}
\end{equation}
where $m_c$ is the central mass, and $m_2$ is the large primary mass.  If the solar mass were higher in the past by a factor $m_f$, then the Hill radius of each planet would have been 
\begin{equation}
   r_H'=m_f^{-4/3}{r_H}
	\label{e:rHratio}
\end{equation}
where ${r_H}$ is the present value.
Thus, a mass losing Sun causes the Hill sphere of each planet to expand; this offers a novel way for the giant planets to capture satellites. This mechanism is similar to the pull-down capture mechanism proposed by~\cite{Heppenheimer:1977Icarus}, which invokes the expansion of the planet's Hill sphere during the gas accretion phase; however, the gas accretion phase is very short-lived, whereas the solar mass loss would occur over a much longer time scale.  As the sun loses mass, its Hill's radius increases as shown in equation~\ref{e:rHratio}.  Objects that are weakly bound to a planet, or are unbound but near the weak stability boundary of the planet may become permanently captured as the Sun's mass decreases~\citep{Astakhov:2004MNRAS}. 
The efficiency of this mechanism is only weakly dependent on planet mass.  We speculate that it may offer an explanation for the observed similarity in the irregular satellite populations of Jupiter, Saturn, Uranus and Neptune~\citep{Jewitt:2005SSR,Sheppard:2006IAUS}.  The long timescale of this process is also an attractive means for evolving irregular satellites into secular resonances~\citep[cf.~][]{Saha:1993Icarus,Nesvorny:2003AJ}.  A detailed examination of these possibilities is beyond the scope of the present paper, but we hope to address them in future work.

We note briefly that the growth of the giant planets as they accreted large amounts of gas during their formation also causes an expansion of the Hill sphere.  This has been invoked in the ``pull-down'' mechanism proposed by~\cite{Heppenheimer:1977Icarus} for the capture of irregular satellites.  In this mechanism, the planetary Hill sphere expands by a much larger magnitude and over a much shorter period of time compared to the solar mass loss mechanism.  Another mechanism for changing planetary Hill spheres is the late orbital migration of the giant planets~\citep{Hahn:1999AJ,Tsiganis:2005Nature} which could also have facilitated the capture of irregular satellites~\citep{Brunini:1995EMP}.

\section{Conclusion}\label{s:Conclusion}
\begin{figure} 
\plotone{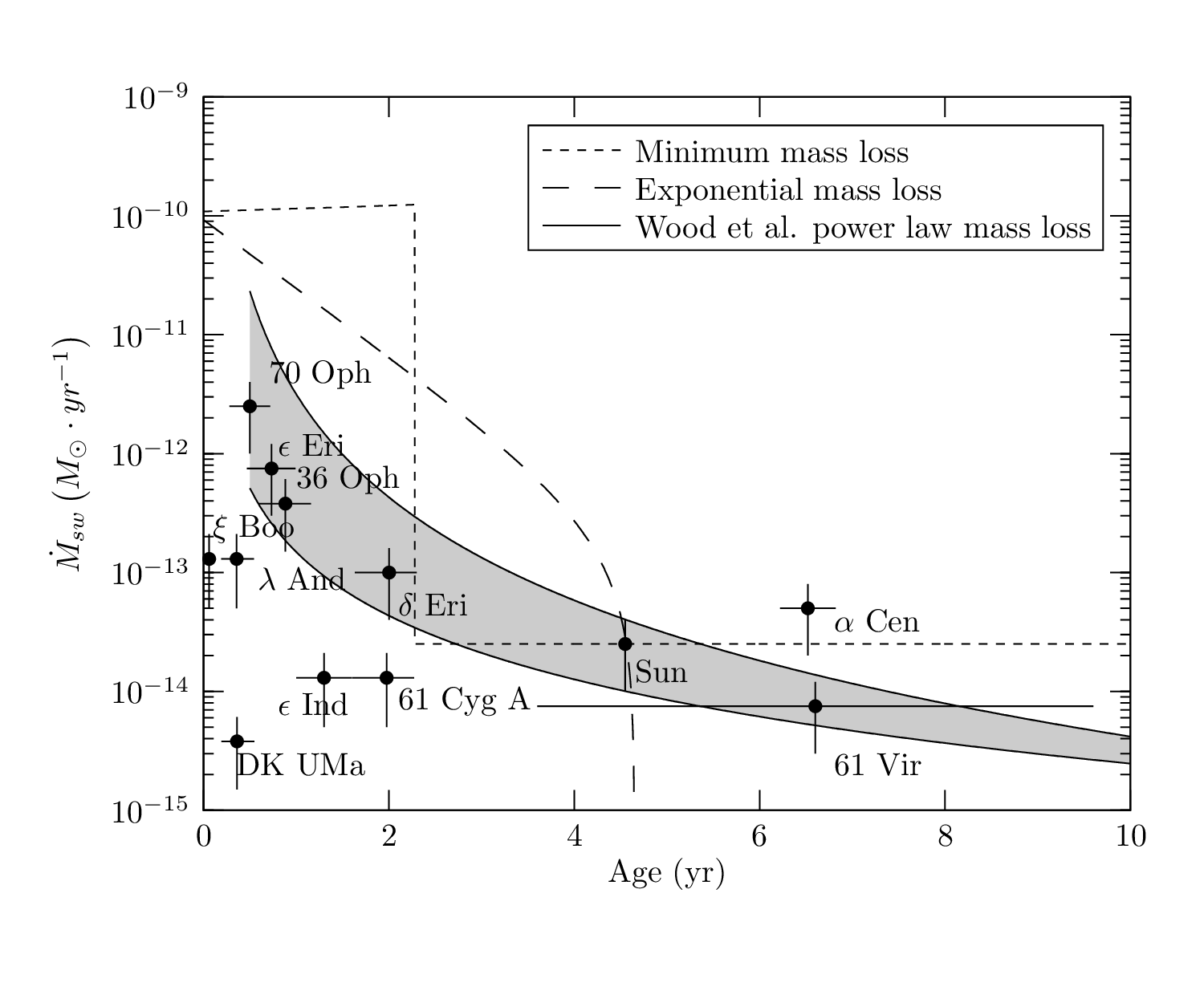}
\caption{Mass loss rates due to stellar wind of the G-K main sequence stars from~\citet{Wood:2005ApJ} are plotted as black circles along with the three different solar mass history models considered.  The vertical error bars are calculated assuming the measured stellar wind mass loss are known to within a factor of two. \label{f:solarmassrate}} 
\end{figure}

We have calculated that the minimal cumulative mass loss of the Sun that would resolve the faint early Sun paradox is $\sim$0.026$\Msun$. Models with a Sun which is up to $\sim$7\% more massive than present are also consistent with helioseimological constraints on solar interior evolution and with geological constraints.  However, an important conclusion of our study is that the solar mass loss history that would resolve the faint early Sun paradox requires the Sun to remain moderately more massive than present for 1--2~Gy in its early history (Figure~\ref{f:solarmassrate}).  

Astronomical evidence (also shown in Figure~\ref{f:solarmassrate}) suggests that both the cumulative mass loss and the timescale for the mass loss (by means of stellar winds) of sun-like stars is significantly short of what is required to resolve the faint early sun paradox.  The cumulative mass loss of solar age sun-like stars on the main sequence, based on the power law mass loss rate time history given by \cite{Wood:2005ApJ}, is $\lesssim$0.3\%, and the mass loss rate declines rapidly in $\sim 10^9\yr$.

Of course, the astronomical data does not directly rule out the possibility that the Sun had a different time history of mass loss than that indicated by the compilation of stellar wind flux estimates of an ensemble sun-like stars of various ages.  We therefore examined the effects of a solar mass loss history on the orbital dynamics in the planetary system, including possible signatures in the geological and climate record.  Most effects are found to be too small to provide unambiguous tests of the early-massive-sun hypothesis.  This hypothesis may offer a novel explanation for the capture of irregular satellites of the outer planets, including the capture into secular resonance for some Jovian irregular satellites.  However, even if the current satellites of the giant planets preserved the signature of a more massive sun, other dynamical events in the early Solar System, such as the accretion and orbital migration of the giant planets may also provide viable explanations.  With the recent accumulation of considerable data on the properties of irregular satellites, it may be possible to test the viability of these alternative mechanisms for the capture of irregular satellites.  This requires more detailed theoretical development of the alternative scenarios for comparison with observations.

\acknowledgments
We thank Alex Pavlov for helpful discussions on Mars climate issues.  RM thanks Jack W.~Szostak for inspiring her interest in this problem.  We are grateful for research support from NASA's Origins of Solar Systems Research Program and from the NASA Astrobiology Institute. 

\bibliographystyle{apj}

\begin{deluxetable}{c c c c} 
\tablecolumns{4}
\tablecaption{Estimated stellar wind mass loss rates of sun-like stars with measured astrospheres, from \cite{Wood:2005ApJ}. \label{t:star_data}} 
\tablewidth{0pc}
\tablehead{
   \colhead{Name} &
   \colhead{X-Ray Age$^\text{\tablenotemark{a}}$} & 
	\colhead{Independent Age} & 
	\colhead{Stellar wind $\dot{M}^\text{\tablenotemark{a}}$} \\
   \colhead{} &
   \colhead{$\left(\times 10^9\yr\right)$} &
   \colhead{$\left(\times 10^9\yr\right)$} &
   \colhead{$\left(\times 10^{-14}\Msun\cdot yr^{-1}\right)$}  
 } 
\startdata  
  Sun 				& \nodata 			& $4.56\pm 0.1^\text{\tablenotemark{b}}$ 	& $2.5\pm 1.5$ \\
  $\alpha$ Cen		& $3.4\pm 0.2$		& $6.52\pm 0.3^\text{\tablenotemark{c}}$ 	& $5.0\pm 3$ \\
  $\epsilon$ Eri 	& $0.72\pm 0.3$ 	& \nodata 											& $75\pm 45$ \\
  61 Cyg A 			& $1.9\pm 0.3$ 	& \nodata											& $1.3\pm 0.8$ \\
  $\epsilon$ Ind 	& $2.3\pm 0.3$		& $1.3\pm  0.3^\text{\tablenotemark{d}}$  & $1.3\pm 0.8$ \\
  36 Oph 			& $0.88\pm 0.3$ 	& \nodata											& $38\pm 23$  \\
  $\lambda$ And 	& $0.36\pm 0.2$ 	& \nodata											& $13\pm 8$  \\ 
  70 Oph 			& $0.90\pm 0.3$	& $0.50\pm 0.22^\text{\tablenotemark{e}}$ & $250\pm 150$ \\
  $\xi$ Boo 		& $0.44\pm 0.2$ 	& $0.06\pm 0.03^\text{\tablenotemark{e}}$ & $13\pm 8$  \\ 
  61 Vir 			& $6.5\pm 0.6$ 	& $6.6\pm 3^\text{\tablenotemark{d}}$ 		& $0.75\pm 0.45$  \\
  $\delta$ Eri 	& $2.0\pm 0.3$ 	& \nodata											& $10.0\pm 6$  \\
  DK Uma  			& $0.36\pm 0.2$ 	& \nodata											& $0.38\pm .23$ \\ 
\enddata
\tablenotetext{a}{\citet{Wood:2005ApJ}}
\tablenotetext{b}{\citet{Dearborn:1991TSIT}}
\tablenotetext{c}{\citet{Eggenberger:2004AA}}
\tablenotetext{d}{\citet{Lachaume:1999AA}}
\tablenotetext{e}{\citet{Barry:1988ApJ}}
\end{deluxetable}
\end{document}